\documentclass{svjour2}                    
\smartqed  
\usepackage{graphicx}
\def\dd{{\rm d}}

\def\_#1{^{}_{#1}}
\def\half{{\textstyle\frac12}}

\def\beq{\begin{equation}}
\def\eeq{\end{equation}}
\def\bea{\begin{eqnarray}}
\def\eea{\end{eqnarray}}
\begin{document}

\title{Solutions of Higher Dimensional Gauss-Bonnet FRW Cosmology}
\author{Keith Andrew \and Brett Bolen \and
        Chad A. Middleton
}


\institute{Keith Andrew \at
              Department of Physics and Astronomy, Western Kentucky University
Bowling Green, KY 42101, U.S.A. \\
              \email{keith.andrew@wku.edu}
           \and
           Brett Bolen \at
              Department of Physics and Astronomy, Western Kentucky University
                Bowling Green, KY 42101, U.S.A.\\
                \email{ brett.bolen@wku.edu} \and
           Chad A. Middleton \at
Department of Physical and Environmental Sciences, Mesa
State College, Grand Junction, CO 81501, U.S.A.\\
\email{chmiddle@mesastate.edu}}
\date{Received: / Accepted: date}

\maketitle

\begin{abstract}
We examine the effect on cosmological evolution of adding a
Gauss-Bonnet term to the standard Einstein-Hilbert action for a
$(1+3) + d$ dimensional Friedman-Robertson-Walker (FRW) metric. By
assuming that the additional dimensions compactify as a power law as
the usual $3$ spatial dimensions expand, we solve the resulting
dynamical equations and find that the solution may be of either de
Sitter or Kasner form depending upon whether the Gauss-Bonnet term
or the Einstein term dominates.

\end{abstract}

\section{Introduction}

Ever since the introduction of extra dimensions by Kaluza and Klein,
the notion that hidden extra dimensions may play a role in the
dynamics of our usual four dimensional spacetime has received
considerable attention  (for a review see Applequist et. al.
\cite{KKBook}), \cite{Ohta}.   However if these extra dimensions exist, experiments
do provide constraints on the maximum current size (of extra dimensions) to be $\sim 100 \, \mu m$.
\cite{BHColliders} An idea that has been extensively studied is that
these extra dimensions were once large but underwent dynamical
compactifaction as the usual three spatial dimensions grew. This has
been studied by Paul and Murkherjee \cite{PaulMuk} and Mohammedi
\cite{Mohammedi} among others.

It is also widely thought that Einstein gravity is only a low energy effective field theory which
requires modification at higher energy. A detailed review of this idea was given by Deser in \cite{Deser}.
One possible modification of the Einstein-Hilbert action comes from adding additional Lovelock terms. \cite{LovelockGRG}
This modification is attractive because it yields second-order, divergence free field equations as one would demand from a
generally covariant theory of gravitation.\cite{Dias} One may formulate Lovelock gravity theory as an expansion in powers of
the curvature to obtain a zeroth-order constant term or cosmological constant, a first-order term that gives the usual scalar
curvature which yields Einstein gravity, and a second-order term that is known as the Gauss-Bonnet term plus higher order terms.

This paper incorporates both of these ideas: compactification of the higher dimensions and the addition of a Gauss-Bonnet term to the action.
We consider a dynamical compactification of a $D$-dimensional manifold to a maximally symmetric manifold of dimension $d$ and an expanding FRW spacetime
of dimension $4$ where we have  modified the Einstein-Hilbert action by including a Gauss-Bonnet (GB) term.\cite{GRGgausbonnet}  This Gauss-Bonnet term
can be interpreted as being a first order correction from string theory or simply a modification of Einstein gravity.

We will choose a power relation between the conformal factors of the three spatial dimensions and the $d$ extra dimensions.
This choice is motivated by the fact when the model is void of Gauss-Bonnet terms, the resulting field equations become exactly that
of 4D FRW cosmology for arbitrary values of $n$ and $d$, after one redefines the coupling and cosmological constant. \cite{Mohammedi}


This paper is organized as follows. In section \ref{framesec} we present a general Einstein-Hilbert plus Gauss-Bonnet action in $D$-dimensions
where the field equations are calculated and the correction to the FRW equation is given. In section \ref{dynamiccompact},  we define an ``effective" pressure which an observer constrained to only the four dimensional volume would observe.   This  arises from the conservation equation, where this defined effective pressure yields an identical expression for the conservation equation to that of a 4D theory.  In section \ref{GenSoln}, we find specific solutions for the $4+d$ dimensional Einstein plus Gauss-Bonnet equations for cases when the Einstein terms or the Gauss-Bonnet terms dominate.  Finally in section \ref{conclusion} we summarize our results.

\section{Framework for Field Equations} \label{framesec}
In this paper we will follow the notation of Paul and Mukherjee\cite{PaulMuk} and Mohammedi
 \cite{Mohammedi} to express the Einstein-Hilbert
action with an extra Gauss-Bonnet term as
\beq\label{action}
     S= \int \dd^D x \sqrt{- g} \left[ R -  \lambda - \epsilon \, \mathcal{G} \right]
\eeq where $\mathcal{G}$ is a Gauss-Bonnet term $(\mathcal{G}\equiv
R_{A B C D} R^{A B C D} - 4 R_{A B} R^{A B} +R^2)$. A variation of
the action (\ref{action}) with respect to $g_{AB}$ produces the equation;
\beq\label{field}
 G_A{}^C+\lambda g_A{}^C+\epsilon\;\mathcal{G}_A{}^C =\frac{1}{\kappa}T_A{}^C
 \eeq
where $\kappa$ is the $4+d$ dimensional coupling constant and the Einstein and Gauss-Bonnet tensors, respectively, are
\bea
 G_A{}^C&=&R_A{}^C -\half g_A{}^C R \\
 \mathcal{G}_A{}^C&=&\frac{1}{2}\left( R_{B D EF} R^{BDEF} -
4 R_{BD} R^{BD} + R^2\right)  \delta_A{}^C \nonumber\\
&&  -\big( 2 R_{BDEA} R^{BDEC} + 2 R R_A{}^C-4 R_D{}^B R_{B A}{}^{D
C}- 4 R_B{}^C R_A{}^B \big) \, .
\eea

Furthermore we will assume that the stress-energy tensor will be
that of a perfect fluid,  thus it is of the form
\beq
T_\mu {}^ \nu= \textrm{diag} \left[ \rho(t), p(t) , p(t), p(t), p_d(t), ...p_d(t)\right]
\eeq
where $p_d(t)$ is the pressure on the higher
dimensional compact manifold.

We will choose a metric ansatz:
\bea
\dd s^2 =& & - \dd t^2 +  \, a^2(t) \left[ \frac{\dd r^2}{1- K r^2} + r^2 \left(\dd \theta^2 + \sin^2
\theta \dd \phi^2 \right) \right] \nonumber\\ & & \quad +\, b^2(t)\gamma_{m n} (y) \dd y^m \dd y^n
\eea
where the extra dimensions are defined to be maximally
symmetric such that the Riemann tensor for $\gamma$ has the form
$R_{abcd} = k (\gamma_{a c} \gamma_{b d} - \gamma_{a d}
\gamma_{b c})$ . In agreement with current observations we will
consider the usual $3$ spatial dimensions to be flat $(K=0)$ and
also demand that the extra dimensions be flat ($k=0$) as well
in agreement with Mohammedi \cite{Mohammedi} that one finds unphysical properties
for $\rho$ and $p$ if $k \ne 0$ .

The metric leads to Riemann tensors of the following form (where both the dimensions are flat)
\bea
R_{a 0 a 0} &=& a \ddot a \, , \quad
R_{a b a b} = a^2 \dot a^2 \, , \quad
R_{m 0 m 0} = b \ddot b \, , \quad \nonumber \\
R_{m a m a} &=&  a \dot a b \dot b \, , \quad
R_{a b a b} =  b^2 \dot b^2
\eea
where $a, b$ are indices which run from $1 ... 3$ and $m,n$ are indices which are in the extra dimensions.    The Ricci Tensor and
Ricci Scalar are
\bea
R_{00} &=&  3 \frac{\ddot a}{a} + d \frac{\ddot b}{b} \, , \quad \quad
R_{a a} =2 \dot a^2 + a \ddot a+d \frac{a \dot a \dot b}{b}  \, , \nonumber\\
R_{m m} &=& 3 \frac{b \dot a \dot b}{a} + (d-1) \dot b^2 + \ddot b b
\eea
\beq
R = 6 \frac{\ddot a}{a} + 2 d \frac{\ddot b}{b} + 6 d \frac{\dot a \dot b}{a b} + 6\frac{\dot a^2}{a^2}  + d (d-1)\frac{\dot b^2}{b^2}
\eeq
where $d$ is the number of extra dimensions.

The full Einstein tensor may be expressed as having an Einstein term $G_\mu {}^\nu$ and a Gauss-Bonnet term $\mathcal{G}_\mu {}^\nu$.
We make the assumption that the extra dimensions compactify as the $3$ spatial dimensions expand as was done by Mohammedi \cite{Mohammedi}
\beq\label{comp}
b(t) \sim\frac{1}{a^n(t)}
\eeq
 where $n>0$ in order to insure that the scale factor of the compact manifold is both dynamical and compactifies as a function of time.
 With this ansatz, the non-zero elements of the Einstein tensor take the form
\bea
G_0 {}^0 &=& - \eta_1 \, \frac{\dot a^2}{a^2} \label{E00} \\
G_a{}^a&=& \eta_2 \,\frac{\ddot a}{a}+ \left(\eta_1-\eta_2\;^2\right)\frac{\dot a^2}{a^2} \label{Eaa} \\
G_m{}^m &=&\frac{1}{d\;n}\left[\left(2\eta_1+3\eta_2\right)\frac{\ddot a}{a}+ \left[2\left(2\eta_1+3\eta_2\right)-d\,n\left(\eta_1+3\eta_2\right)\right]\frac{\dot a^2}{a^2}\right]
  \label{Emm} \, .
\eea
where we defined the coefficients
\bea
\eta_1&=&\frac{1}{2}\left[6+d \, n(d \, n-n-6)\right]\nonumber\\
\eta_2&=&\left(d\;n-2 \right)
\eea
Note that when $D=4$ (or $d=0$) the equations become the well known FRW equations in four dimensions.
In $4+d$ dimensions, the Gauss-Bonnet terms become
\bea
\mathcal{G}_0 {}^0 &=&  \xi_1 \frac{\dot a^4}{a^4}\\
\mathcal{G}_a {}^a &=& \frac{1}{3}\left(\xi_1-dn\;\xi_3\right)\frac{\dot a^4}{a^4} + \xi_3 \, \frac{\ddot a\dot a^2}{a^3} \\
\mathcal{G}_m {}^m &=&-\left(\xi_1+\xi_3\right)\frac{\dot a^4}{a^4}  + \frac{1}{d\;n}\left(3\;\xi_3+4\;\xi_1\right)\frac{\ddot a \dot a^2}{a^3}
\eea
where we defined the coefficients
\bea
 \xi_1&=&-d \, n \left[(d-1)n\left[\frac{1}{2}(d-2)n\left[(d-3)n-12\right]+18\right]-12\right]\nonumber\\
 \xi_3&=&\quad - 2d \, n\Bigg[ (d-1)n\left[ (d-2)n-6 \right]+6\Bigg]
 \eea
where the constants $\eta_i$ and $\xi_i$ depend upon the values of $n$ and $d$ as defined
above. Note that if we force the number of extra dimensions  to zero $(d \rightarrow 0)$ then the Gauss-Bonnet terms vanish for $\mathcal{G}_0 {}^0$
and $\mathcal{G}_a {}^a$ as one would expect in four dimensions.

\section{Effective Pressure and the Field Equations} \label{dynamiccompact}
 For the case of both spaces having flat curvature, the
$D$-dimensional  Friedmann-Robertson-Walker (FRW) equations (\ref{field}) take the form
\bea
\frac{\rho}{2 \kappa} &=&\eta_1\frac{\dot a^2}{a^2} + \epsilon \, \xi_1 \, \frac{\dot a^4}{a^4}  \label{density1}\\
\frac{p}{2 \kappa} &=&\left[\eta_2 \,\frac{\ddot a}{a}+ \left(\eta_1-\eta_2\;^2\right)\frac{\dot a^2}{a^2}\right]+ \epsilon\left[\frac{1}{3}\left(\xi_1-dn\;\xi_3\right)\frac{\dot a^4}{a^4} + \xi_3 \, \frac{\ddot a\dot a^2}{a^3}\right]  \label{presure1} \\
\frac{p_d}{2 \kappa} &=&\frac{1}{d\;n}\left[\left(2\eta_1+3\eta_2\right)\frac{\ddot a}{a}+\left[2\left(2\eta_1+3\eta_2\right)-d\,n\left(\eta_1+3\eta_2\right)\right]\frac{\dot a^2}{a^2}\right]\nonumber\\
&-& \epsilon \left[\left(\xi_1+\xi_3\right)\frac{\dot a^4}{a^4}  - \frac{1}{d\;n}\left(4\;\xi_1+3\;\xi_3\right)\frac{\ddot a \dot a^2}{a^3}\right]\,\label{presD}
\eea
where we have set $\lambda=0$.  Together with these Einstein
equations, we also demand that the conservation equation hold for
the stress-energy tensor $\left( \nabla_\mu T^\mu{}_\nu=0 \right)$
or
\beq\label{cons}
\left\{\frac{\dd}{\dd t} (a^3 \rho) + p \, \frac{\dd}{\dd t} (a^3) \right\} + d \, a^3 \, \frac{\dot b}{b} \left( \rho +  p_d \right) = 0
\eeq
Using the assumption that
$b=1/a^n$, this becomes
\beq \frac{\dd}{\dd t} (a^3 \rho) +\tilde{p} \, \frac{\dd}{\dd t} (a^3)= 0 \eeq
which by simple algebra
may be written in the more familiar form
\beq\label{EoS}
\dot{\rho}+3\frac{\dot{a}}{a}(\rho+\tilde{p})=0 \, . \eeq
Note that
this is simply a statement that $\dd E = - P \, \dd V$ where we have
defined an ``effective" pressure $\left( \tilde p \right)$
 \cite{Mohammedi} which an observer
constrained to exist only upon the ``usual" $3$ spatial dimensions
would see as
 \beq\label{effpres}
\tilde{p}=p-\frac{1}{3}d\,n\;(\rho+p_d).
 \eeq
As was pointed out by Mohammedi \cite{Mohammedi}, this effective pressure can be negative for positive values of $\rho,\;p,\mbox{and}\;p_d$.
The effective pressure can be easily computed from the $d$-dimensional FRW
equations (\ref{density1})-(\ref{presD}) and is given by the
relation \bea\label{presT}
 \frac{\tilde{p}}{2\kappa}&=&-\frac{1}{3} \eta_1 \left(2 \frac{\ddot a}{a} +
\frac{\dot a^2}{a^2}\right) -\frac{1}{3}\epsilon\xi_1 \left( 4\, \frac{\ddot a\dot a^2}{a^3}-\frac{\dot a^4}{a^4} \right)
\eea
 Note that the above field equations (\ref{density1})-(\ref{presD}), (\ref{presT}) become the same as \cite{Mohammedi} in the limit where
$\epsilon \rightarrow 0$ (i.e. no Gauss-Bonnet terms).  By redefining the coupling and cosmological constant in eqns. (\ref{density1}) and (\ref{presT}), one recovers standard 4-D FRW cosmology for arbitrary values of $n$ and $d$ in the $\epsilon \rightarrow 0$ limit.

\section{The General Solutions of the GB FRW Field Equations}\label{GenSoln}
The effective $D$-dimensional FRW equations and the conservation equation now read
\bea
0&=&\dot{\rho}+3\frac{\dot{a}}{a}(\rho+\tilde{p})\label{EoS1}\\
\frac{\rho}{2 \kappa} &=&\eta_1\frac{\dot a^2}{a^2} + \epsilon \, \xi_1 \, \frac{\dot a^4}{a^4} \label{density2}\\
 \frac{\tilde{p}}{2\kappa}&=&-\frac{1}{3} \eta_1 \left(2 \frac{\ddot a}{a} +
\frac{\dot a^2}{a^2}\right) -\frac{1}{3}\epsilon\;\xi_1 \left( 4\, \frac{\ddot a}{a}-\frac{\dot a^2}{a^2} \right)\frac{\dot a^2}{a^2} \label{presT2}\\
\frac{p_d}{2 \kappa} &=&\frac{1}{d\;n}\left[\left(2\eta_1+3\eta_2\right)\left(\frac{\ddot a}{a}+2\frac{\dot a^2}{a^2}\right)-d\,n\left(\eta_1+3\eta_2\right)\frac{\dot a^2}{a^2}\right]\nonumber\\
&+& \epsilon\; \frac{1}{d\;n}\left[3\left(\xi_1+\xi_3\right)\left(\frac{\ddot a}{a}-\frac{1}{3}d\;n\frac{\dot a^2}{a^2}\right)  +\xi_1\frac{\ddot a}{a}\right]\frac{\dot a^2}{a^2} \,\label{presD2}
\eea
One can easily show that (\ref{EoS1}) is trivially satisfied when (\ref{density2})
and (\ref{presT2}) are employed.
This is exactly like that of 4D FRW cosmology where the variables are left underdetermined by the field equations and one usually chooses an equation of state (EoS) of the form
\beq\label{w}
\tilde{p}=w\;\rho
\eeq
to proceed.  Notice that the parameter $w$ can in general be time-dependent.   If $w$ is in fact constant, (\ref{EoS1}) can be integrated yielding
\beq\label{rhoa}
\rho(t)=\rho_0 a^{-3(1+w)}
\eeq

Using (\ref{w}) and eliminating $\rho$ and $\tilde{p}$ from the field equations, we obtain an expression of the form
\beq\label{adota}
(1+w)\left(\eta_1+\epsilon\xi_1\frac{\dot a^2}{a^2}\right)\frac{\dot a^2}{a^2}=-\frac{2}{3}\left(\eta_1+2\epsilon\xi_1\frac{\dot a^2}{a^2}\right)\frac{d}{dt}\left(\frac{\dot a}{a}\right)
\eeq
 The behavior of the scale factor is determined by this expression.  As is obvious from (\ref{adota}), the scale factor is dependent on the value
 of the parameter $w$ as one would expect.   In the following subsections, we will examine the solutions of this equation for cases when parameter
 is uniquely $w=-1$ and the general case, when $w\neq -1$.

\subsection{Solutions for $w=-1$}

If $w=-1$, the equation of state (\ref{adota}) reduces to
\beq\label{adota1}
0=\left(\eta_1+2\epsilon\xi_1\frac{\dot a^2}{a^2}\right)\frac{d}{dt}\left(\frac{\dot a}{a}\right)
\eeq
This equation has two solutions depending on which bracket is equated to zero.  In either case, we find a de Sitter-type solution given by the expressions
\beq
a(t)=a_0 \;e^{H t}\label{deS2}
\eeq
where $H$ is the Hubble constant which is obtained from the remaining field equations.
Plugging $a(t)$ into (\ref{density2}) yields a value for the Hubble parameter of the form
\beq\label{H}
H\;^2=\left( \frac{\dot{a}}{a} \right)^2 =-\frac{\eta_1}{2\epsilon\xi_1}\left[1\pm\sqrt{1+\frac{4\epsilon\xi_1}{\eta_1^2}\cdot\frac{\rho}{2\kappa}}\;\right] \, .
\eeq
Notice that when the vacuum-energy density is chosen such that the radical vanishes, we find a result corresponding to what one obtains when the
first bracket is set equal to zero.  Hence, the first solution is simply a special case of the second solution.

In the small epsilon limit, when the Gauss-Bonnet contribution
offers a small correction to the Einstein-Hilbert term, we find
\beq
H^2\simeq \frac{\rho}{2\kappa\eta_1}\left[1-\epsilon\;\frac{\xi_1}{\eta_1^2}\frac{\rho}{2\kappa}\right]
\eeq
to $\cal{O}(\epsilon)$.  Note that we kept only the negative root of
(\ref{H}) so that the value of the Hubble constant approaches that
of 4D FRW cosmology in this small $\epsilon$ limit.

 In the large epsilon limit, when the Gauss-Bonnet term dominates over the Einstein term, we find a value for the Hubble constant of the form
\beq\label{lgepw=-1}
H^2\simeq\pm\frac{1}{\beta}\sqrt{\frac{\rho}{2\kappa\eta_1}}-\frac{1}{2\beta^2}
\eeq
where we defined the parameter
\beq\label{beta}
\beta=\sqrt{\frac{\epsilon\xi_1}{\eta_1}}
\eeq
and kept terms to $\mathcal{O}(1/\epsilon)$.

We can eliminate the expansion factor and obtain a higher dimensional equation relating both $p_d$ and $\rho$.
This  EoS-like expression takes the form
 \beq
(p_d-\psi_0\rho)+\frac{\epsilon}{2}\frac{\sigma}{(\psi_0-\chi_0)}(p_d-\chi_0\rho)^2= 0\label{dEoS}
\eeq
where we defined the constants
\bea\label{wpar}
\psi_0&=&-\frac{1}{\eta_1}\left[\left(1-\frac{3}{d\;n}\right)(2\eta_1+3\eta_2)-\eta_1\right]\nonumber\\
\chi_0&=&-\frac{1}{\xi_1}\left[\left(1-\frac{3}{d\;n}\right)(\xi_1+\xi_3)-\frac{1}{d\;n}\;\xi_1\right] \, .
\eea
Notice that (\ref{dEoS}) is inherently {\it non-linear} as one might expect from an exact solution to Gauss-Bonnet FRW cosmology.
Also notice that when $d\,n=3$, which corresponds to a constant volume $d+4$ dimensional Universe, (\ref{wpar}) dramatically simplifies
yielding the fixed parameters $\psi_0=1$ and $\chi_0=1/3$.

\subsection{Solution for $w\neq-1$}

In this subsection, we investigate the behavior of the scale factor for the general case when $w$ is left as a free parameter.
By integrating the equation of state, eqn. (\ref{adota})
\beq\label{adota2}
\beta\frac{\dot a}{a}=\tan\left[\frac{1}{\beta}\left(\frac{1}{\dot a/a}-\frac{3}{2}(1+w)t\right)\right]
\eeq
where we again defined $\beta= \sqrt{\epsilon \xi_1 / \eta_1}$. In this section,
we can explore this behavior by investigating
(\ref{adota2}) when the Gauss Bonnet term is small and when it dominates over the Einsteinian term.

\subsubsection{Small $epsilon$ regime}

 In the small $\epsilon$ regime, one can expand the equation of state (\ref{adota2}) by taking the $\arctan$ of each side and taking the expansion
for small $x$.
Keeping only the lowest order contribution in the expansion, it becomes
\beq\label{quad}
\beta^2\;\frac{\dot a^2}{a^2}+\frac{3}{2}(1+w)\;t\;\frac{\dot a}{a}-1\simeq 0 \, .
\eeq
Solving for the Hubble parameter, we obtain
\beq
\frac{\dot a}{a}\simeq-\frac{3}{4\beta^2}(1+w)\;t\left[1\pm\sqrt{1+\left[\frac{4\beta}{3(1+w)t}\right]^2}\;\right] \, .
\eeq
Again, expanding the square root in the small $\epsilon\;(\sim\beta^2)$ limit and keeping the first order $\epsilon$ contribution, one obtains
\beq\label{smep}
H(t)=\frac{\dot a}{a}\simeq\frac{2}{3(1+w)}\frac{1}{t}\left[1-\epsilon\;\frac{\xi_1}{\eta_1}\left[\frac{2}{3(1+w)}\frac{1}{t}\right]^2\right]
\eeq
where we again kept only the negative root in order to arrive at 4D FRW cosmology in this small $\epsilon$ limit.
 Using this result,  we integrate (\ref{smep}) and obtain a value for the scale factor of the form
 \beq\label{a}
a(t) \simeq\mu\; t\;^{2/3(1+w)}\left[1+\epsilon\;\frac{\xi_1}{2\eta_1}\left[\frac{2}{3(1+w)}\right]^3\frac{1}{t^2}\right]
\eeq
to first order in $\epsilon$.  Note that in the large time limit $(t\rightarrow\infty)$ the scale factor tends to its zeroth-order value.
Hence, we see that the Gauss-Bonnet contribution becomes vanishingly insignificant for late cosmological times as one would expect.

As a consistency check,  we insert (\ref{rhoa}), (\ref{smep}), and (\ref{a}) into (\ref{density2}) and find that the time dependence does in fact vanish to first order in $\epsilon$.  We obtain a relation on the coefficients given by
\beq\label{mu}
\mu=\left[\left[\frac{3}{2}(1+w)\right]^2\frac{\rho_0}{2\kappa\eta_1}\right]^{1/3(1+w)}
\eeq
which agrees with the value one obtains from 4D FRW cosmology.
\subsubsection{Large $\epsilon$ regime}
Interestingly, because the first order term in the the expansions of $\tan(x)$ and $\arctan(x)$ are both linear in $x$;
 the expansion for when the Gauss-Bonnet term dominates (large $\epsilon$ or equivalently when $t \ll \beta$),(\ref{adota2}) is the same as for small $\epsilon$ (\ref{quad})
\beq\label{adota3}
\beta\frac{\dot a}{a}= \frac{1}{\beta}\left(\frac{1}{\dot a/a}-\frac{3}{2}(1+w)t\right) \, .
\eeq
The difference between the large and small $\epsilon$ cases is that in the large $\epsilon$ case
one does not expand the radical as was done in (\ref{smep}), instead one may simply express the equation in terms of $H$ as
\beq
H^2 + \left[ \frac{3 t}{2 \beta^2} (1 +w) \right] H - \frac{1}{\beta^2} =0 \, .
\eeq
If solves this quadratic and keeps the terms large in $1/ \epsilon$ (the terms that dominate when GB is large), one obtains
\beq
a(t) \approx a_0 \textrm{Exp} \left[\frac{t}{\beta}  - \frac{3 \, t^2}{8 \beta^2} (1+w) + \mathcal{O} \left(\frac{t^3}{\beta^3} \right) \right] \, .
\eeq
Note that when the Gauss Bonnet terms dominate  the form of $a(t)$ is de Sitter like.

\section{Conclusion} \label{conclusion}

In conclusion, we have studied the Friedman equations for an Einstein plus
Gauss-Bonnet action in $4 + d$ dimensions.  We furthermore demanded that the extra
dimensions compactify as $a(t) \sim b(t)^{-n}$ where $n>0$.  In Section \ref{GenSoln}, we solved the field equations for cases when $w=-1$ and $w \ne -1$.
We found solutions for both of these cases when the Gauss-Bonnet term dominates the Einstein term early time cosmology and when the
Gauss-Bonnet term is subdominant to the Einstein term (late time cosmology).  We have found that when Einstein gravity dominates, the conformal factor is
 a Kasner solution with a small Gauss-Bonnet correction.  However, when the Gauss-Bonnet term dominates, a de Sitter type solution is obtained indicating
 that the Gauss-Bonnet term gives rise to nontrivial corrections of the scale factor $a(t)$.  One may conclude from this analysis that if the Gauss-Bonnet term is
dominant, the compactifying extra dimensions can be thought of as playing the role of a ``cosmological constant''  forcing $a(t)$ to behave as a de Sitter solution even though no such constant is
present in the action.


The natural question which arises from this paper is when does this model of compactifing internal dimensions with extra Gauss-Bonnet terms physically
representative of our universe. Without further modification of the paper, it seems to be applicable only to early universe inflation as our model tends
towards a Kasner solution in late time.  In order to agree with current cosmological observations, the model should allow the extra dimensions to compactify
much slower then the three spatial dimensions grow to be in order to agree with the current limits on the running of the coupling constants.

This paper leads to several intriguing questions.  First, does this same effect of compactifing extra dimensions leading to an effective cosmological
constant exist in alternate models such as brane models like that of Randall-Sundrum \cite{randall} or Narain et al \cite{Witten}.  If so, then could this be used to generate an inflationary cosmology without the use of a scalar field \cite{Neupane}.  A second and central issue to explore would be to relax the explicit relation between $a(t)$ and $b(t)$, could one then solve the resulting field equations?  A possibility we are examining seems to require a numerical study.

 \end{document}